\documentclass[12pt]{aa}
\usepackage[english]{babel}
\usepackage{graphicx}

\onecolumn
\begin{document}

\title{New in the optical spectrum and kinematic state of the atmosphere of the variable V1027\,Cyg (= IRAS\,20004+2955)}
\author{V.G. Klochkova, V.E. Panchuk, and N.S. Tavolganskaya}
\institute{Special Astrophysical Observatory RAS, Nizhnij Arkhyz, 369167 Russia}

\date{\today} 

\abstract {Based on  high-resolution spectroscopy performed with the NES echelle spectrograph of 
the 6-m telescope, we have studied the peculiarities of the spectrum and the velocity field in the 
atmosphere and envelope of the cool supergiant V1027\,Cyg, the optical counterpart of the infrared source
IRAS\,20004+2955. For the first time, a splitting of the cores of strong absorptions of metals and their ions (Si\,II, Ni\,I, Ti\,I,
Ti\,II, Sc\,II, Cr\,I, Fe\,I, Fe\,II, Ba\,II) has been detected in the stellar spectrum. 
The broad profile of these lines contains a stable weak emission in the core whose position may be 
considered as the systematic velocity Vsys\,=\,5.5\,km/s. Small radial velocity variations with an 
amplitude of 5--6\,km/s due to pulsations have been revealed by symmetric low- and moderate-intensity absorptions. 
A longwavelength  shift of the H$\alpha$ profile due to line core distortion is observed in the stellar spectrum. 
Numerous weak CN molecular lines and the K\,I~7696\,\AA\ line with a P\,Cyg profile have been identified 
in the red spectral region. The coincidence of the radial velocities measured from symmetric metal 
absorptions and CN lines suggests that the CN spectrum is formed in the stellar atmosphere. We have also 
identified numerous diffuse interstellar bands (DIBs) whose positions in the spectrum, Vr(DIBs)\,=$-$12.0\,km/s, 
correspond to the velocity of the interstellar medium in the Local arm of the Galaxy.  \newline
 }

\authorrunning{\it Klochkova et al.}
\titlerunning{\it The optical spectrum of the variable V1027\,Cyg}

\maketitle

\section{Introduction}

The cool supergiant V1027\,Cyg of MK spectral type Sp\,=\,G7\,Ia (Keenan and McNeil, 1989) is associated  
with the infrared source IRAS\,20004+2955. Volk and Kwok~(1989) modeled the peculiar, very flat,
IR spectrum of IRAS\,20004+2955 and concluded that the object is observed at the very beginning of
the evolutionary transition from the asymptotic giant branch (AGB) to a planetary nebula. Kwok~(1993)
entered this star in his widely known list of protoplanetary nebula (PPN) candidates, stars in transition to
the planetary nebular stage. However, V1027\,Cyg has remained poorly studied until the present time;
there is no consensus on the evolutionary status of this star.

Based on UBV photometry and low-resolution spectroscopy, Arkhipova et al.~(1992) classified this
star as a semiregular variable with an amplitude of brightness variations increasing from V to U from 0.4
to 0.8~mag. These authors concluded that interstellar extinction is responsible for the bulk of the reddening in 
V1027\,Cyg, while the pattern of brightness and color variability suggested that its variability is
caused by pulsations. Subsequently, having measured the radial velocities Vr with a correlation spectrometer, 
Arkhipova et al.~(1997) detected a radial velocity variability in the range $-10\div +20$\,km/s and pointed 
out a strengthening of Ba\,II lines. All the values of Vr from Hrivnak and Wenxian~(2000) also fit into this 
velocity range.

Based on the spectra taken with the echelle spectrograph of the 6-m telescope, Klochkova et al.~(2000a) 
determined the main parameters of V1027\,Cyg (effective temperature Teff\,=\,5000\,K, surface gravity
log\,g\,=\,1.0), its metallicity ${\rm [Fe/H]}_{\odot}$\,=$-$0.2, and the abundances of 16 elements in the stellar 
atmosphere by the model atmosphere method. The nearly solar metallicity in combination with the low radial velocity
of the star suggest that V1027\,Cyg belongs to the Galactic disk population. Taranova et al.~(2009)
performed long-term JHKLM photometry for yellow supergiants and detected a small-amplitude ($\le 0.25$\,mag) 
IR brightness variability with a probable period of about 237~days in V1027\,Cyg. Using the same observations, 
Bogdanov and Taranova~(2009) computed the model of a spherical circumstellar envelope and estimated the mass loss rate to be
1.3$\times 10^{-5}\mathcal{M}_{\odot}/$yr.

The radial velocity variability of V1027\,Cyg and the evidence for the peculiarity of its spectrum found
by Klochkova et al.~(2000a) serve to us as a stimulus to continue the observations of V1027\,Cyg.
In this paper we present the results of our optical spectroscopy for this star performed with the 6-m telescope 
over several nights in 2013--2015. The goal of these observations is a detailed study of the stellar
spectrum and its possible variability as well as the kinematic state of the atmosphere of V1027\,Cyg, its
circumstellar envelope, and the interstellar medium toward this star with Galactic coordinates l\,=\,67.4$\degr$
and b\,=$-0.3\degr$.

\begin{table} 
\caption{Times of observations of V1027\,Cyg with the NES spectrograph of the 6-m telescope and the recorded wavelength 
         range $\Delta\lambda$  in \AA{}.  The heliocentric radial velocities Vr measured from numerous symmetric absorptions 
         (abs), H$\alpha$ cores, and DIBs are given. The number of lines used to determine Vr is given in parentheses.}
\begin{tabular}{r| c| c | l |c | c}   
\hline
\small Date & \small JD    & $\Delta\lambda$, &\multicolumn{3}{c}{\small  Vr, km/s }\\   
\cline{4-6}
             &  $-2450000$  &  \AA{}                 &\hspace{1cm} abs& H$\alpha$   & DIBs \\ 
\hline               
13.10.2013&6579.32 &4000--6980 &$-4.1\pm 0.1$\,(332) & +11.9  &$-11.4\pm0.4$\,(15)\\ 
28.04.2015&7141.14 &5400--8470 &$+4.5\pm 0.1$\,(264) & +13.3  &$-12.9\pm0.5$\,(13)\\ 
          &&       &$+4.0\pm 0.3$\,(24, CN)          &        &                    \\     
 3.09.2015&7269.36 &4000--6980 & $+9.1\pm 0.1$\,(541)& +7.9   &$-11.7\pm0.4$\,(10) \\ 
\hline
\end{tabular}
\label{Spectra}
\end{table}

Our observational data are briefly described in the next section. Subsequently, we provide information
about the peculiarities of the profiles for the spectral features detected from high-resolution spectra,
analyze them, and discuss our results. Our brief conclusions are presented in the last section.

\section{Observational data}	

We took the spectra of V1027\,Cyg at the Nasmyth focus of the 6-m telescope with the NES echelle
spectrograph (Panchuk et al. 2007, 2009). In combination with an image slicer (Panchuk et al. 2003),
the NES spectrograph provides a spectral resolution R$\approx$60\,000 in a wide wavelength range. 
A 2048$\times$4096-pixel CCD array has been used at the NES spectrograph since 2011,  which has allowed 
the simultaneously recorded spectral range to be extended considerably. The times of observations of V1027\,Cyg
with the 6-m telescope and the recorded spectral range are given in Table\,\ref{Spectra}.
In addition, for comparison, we invoke the stellar spectra taken during several observational sets in
1997 with the NES and PFES echelle spectrographs and used in Klochkova et al.~(2000a). The NES
spectrograph in its first modifcation in combination with a 1k$\times$1k CCD array provided a resolution
R$\approx$35\,000. The PFES spectrograph designed for the observations of faint stars with a resolution 
R$\approx$15\,000 (Panchuk et al.~1997) is mounted at the prime focus of the 6-m telescope.

The details of our spectrophotometric and positional measurements in the spectra were described
in previously published papers; the corresponding references to them were given by Klochkova~(2014).
Note that applying the image slicer required a significant modiﬁcation of the standard ECHELLE procedures 
of the MIDAS software package. The cosmicray particle hits were removed by median averaging of two spectra 
taken successively one after another. The wavelength calibration was performed using the
spectra of a hollow-cathode Th--Ar lamp. The data were extracted from two-dimensional echelle spectra
with the software package described by Yushkin and Klochkova (2005). The latest version of the DECH code 
(Galazutdinov 1992) was used to reduce the extracted spectra. In particular, it allows the radial
velocities to be measured from individual features of complex lines typical of the spectra of the investigated
stars. The positional zero point of each spectrogram was determined in a standard way, by referencing to
the positions of ionospheric night-sky emissions and telluric absorptions, which are observed against the 
background of the object spectrum. The accuracy of measuring the velocity from {\bf one} line in
the NES spectra is $\approx$1.0\,km/s.

\section{Main results}

\subsection*{Previously unknown peculiarities of the spectrum}

\begin{figure}
\includegraphics[angle=0,width=0.5\columnwidth,bb=20 30 570 780,clip]{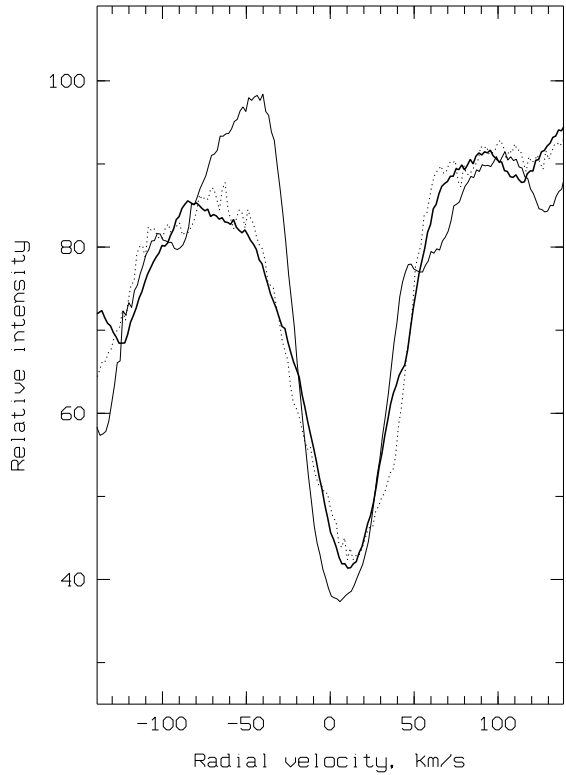}  
\caption{H$\alpha$ profiles in the spectrum of V1027\,Cyg for different dates of observations: October\,13,\,2013 (thick solid curve), 
        April\,28,\,2015 (dotted curve), and September\,3,\,2015 (thin solid curve).}        
\label{Halpha-3dat}
\end{figure}

On the whole, the high-resolution optical spectrum of V1027\,Cyg corresponds to its MK classificaion 
G7\,Ia (Keenan and McNeil, 1989). The absence of a visible emission in H$\alpha$ is quite unexpected for 
a pulsating variable (see Fig.\,\ref{Halpha-3dat}). An emission in H$\alpha$ was also absent in the earlier spectra of this star
taken with the 6-m telescope and used by Klochkova et al.~(2000a) to determine the atmospheric chemical
composition of this star. A strong (above the continuum level) and time-variable emission is a typical signature 
of long-period variables (LPVs). The well studied LPV star R\,Sct, in the optical spectra of
which the intensity of the emission in H$\alpha$ at some phases exceeds the continuum level manyfold (Lebre and Gillet~1991; 
Kipper and Klochkova 2013), can serve as an example. An equally powerful emission in H$\alpha$ was recorded by Klochkova et al.~(2006) 
in the spectrum of the AGB star identified with the IR source IRAS\,20508+2011. Over the four-year period of observations 
of this object with the 6-m telescope the absorption--emission H$\alpha$ profile changed from a bell shaped emission with 
a weak absorption in the core to a double--peaked emission with a central absorption lying below the continuum.

\begin{figure}[ht!] 
\includegraphics[angle=0,width=0.8\columnwidth,height=0.6\columnwidth,bb=20 30 570 780,clip]{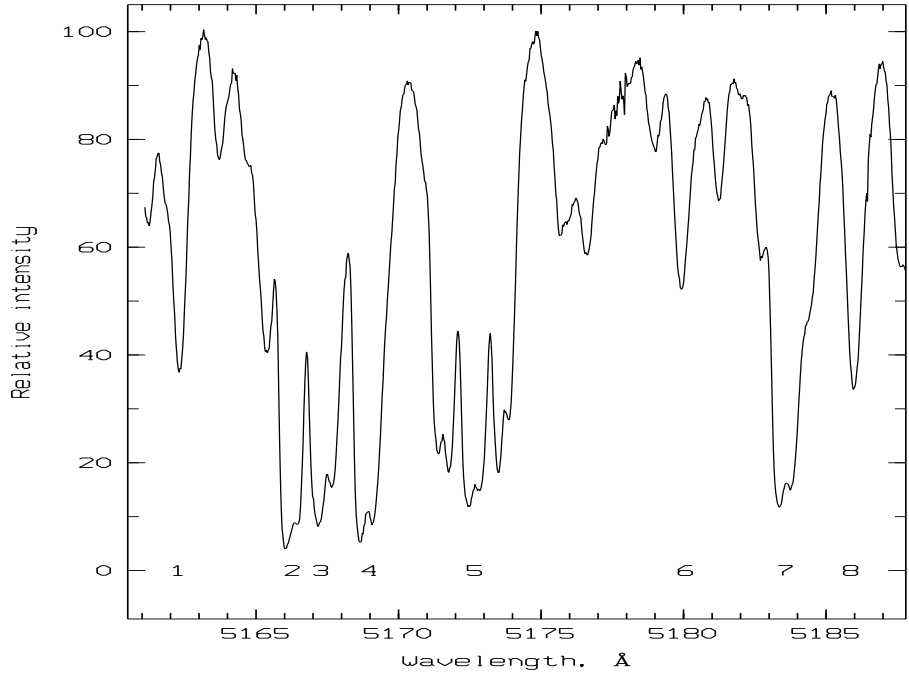}   
\caption{Fragment of the September 3, 2015 spectrum of V1027\,Cyg in the spectral region saturated by absorptions with split
        cores. The identified main absorptions are indicated: Fe\,I~5162.3\,\AA{}~(1), Fe\,I~5166.3\,\AA{}~(2), Mg\,I~5167.3\,\AA{}~(3), 
        Fe\,II~5169.0\,\AA{}~(4), Mg\,I~5172.7\,\AA{}~(5), Fe\,I~5180.1\,\AA{}~(6), Mg\,I~5183.6\,\AA{}~(7), and Ti\,II~5185.9\,\AA{}~(8).}
\label{Split}
\end{figure}

\begin{figure}[ht!] 
\includegraphics[angle=0,width=0.49\columnwidth,height=0.6\columnwidth,bb=20 30 570 780,clip]{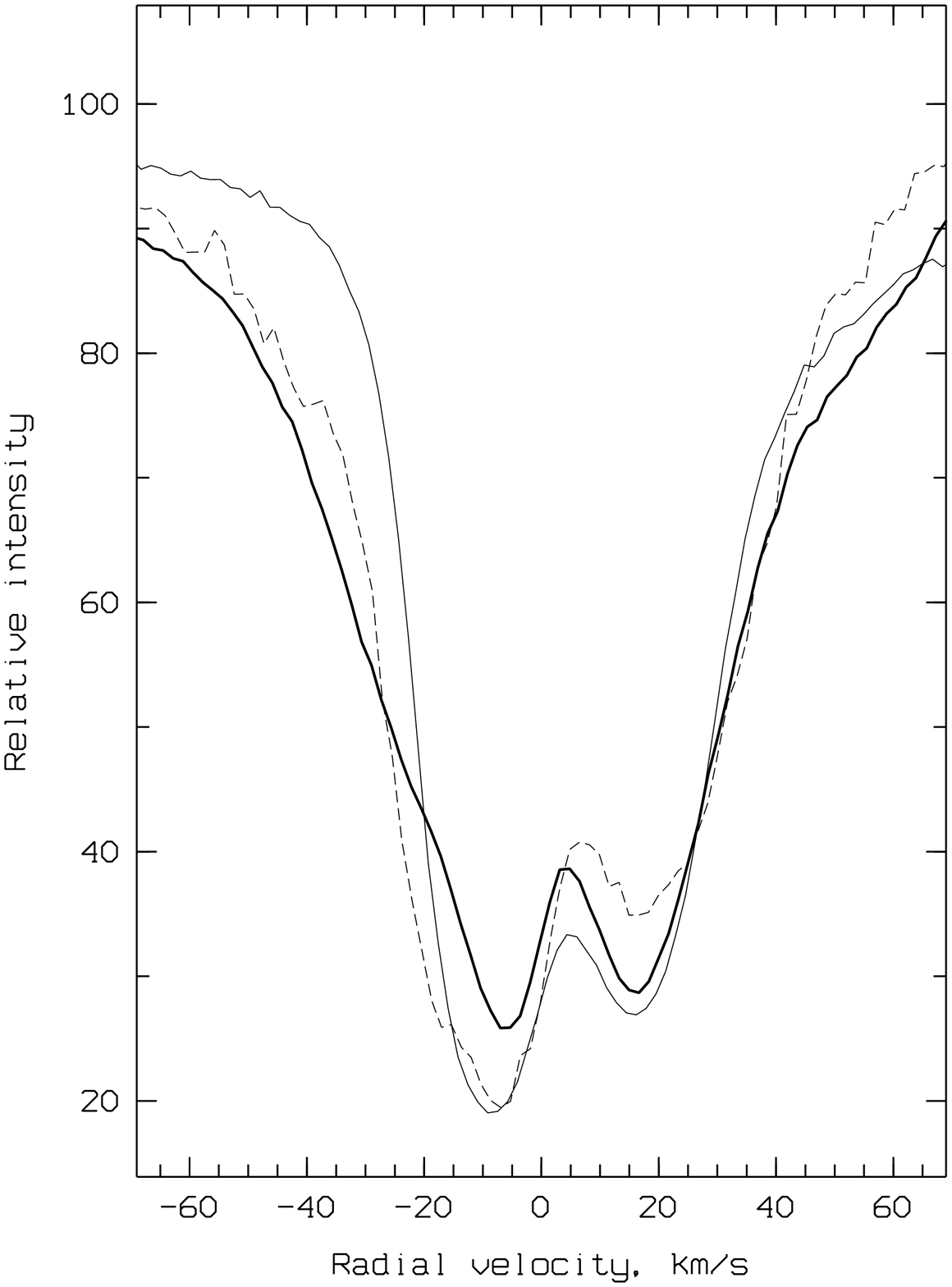}
\includegraphics[angle=0,width=0.49\columnwidth,height=0.6\columnwidth,bb=20 30 570 780,clip]{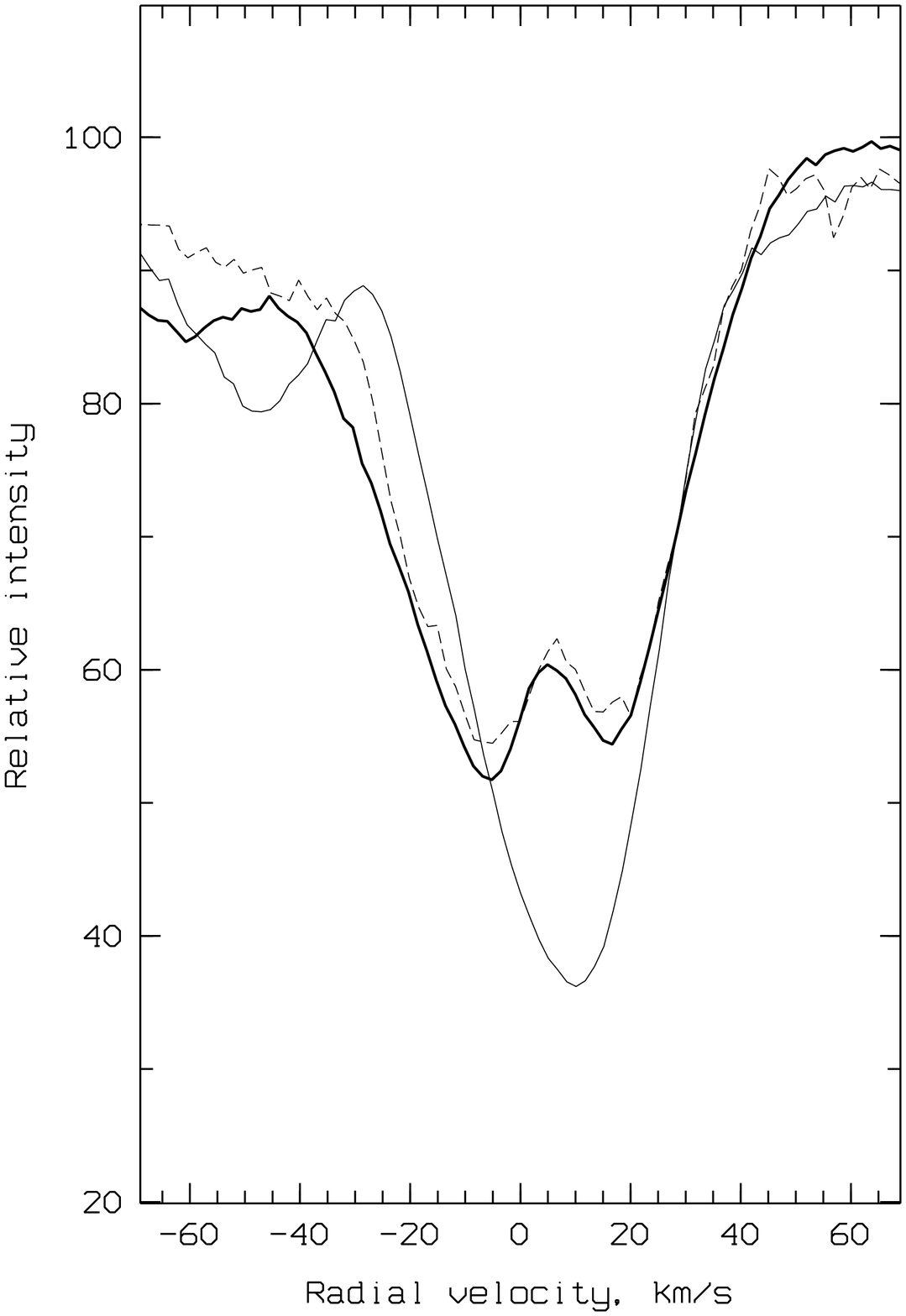}
\caption{Split Ba\,II~6141.42\,\AA{} (left) and Fe\,I~6393.6\,\AA{} (right) line profiles in the spectra of V1027\,Cyg for three dates of
observations: October~13, 2013 (thick solid curve), April~28, 2015 (dashed curve), and September~3, 2015 (thin solid curve).}
\label{BaFe}
\end{figure}

However, in the absence of an emission in H$\alpha$, we detected a number of other, subtler peculiarities in the
spectrum of V1027\,Cyg due to the high quality of our observational data. We saw a new peculiarity of
strong absorptions: the lines with a low lower-level excitation potential ($\chi_{low} < 1.5$\,eV) have a complex
profile including two absorptions with close intensities separated by an emission peak. To illustrate this peculiarity, 
Fig.\,\ref{Split} presents a fragment of the spectrum containing the strong and split Mg\,I, Fe\,I, Fe\,II, and Ti\,II absorptions. 
In addition, it emerged that the profiles of such strong absorptions are variable in time. In Fig.\,\ref{BaFe}, where the profiles of the 
two strong Ba\,II~6141\,\AA{} and Fe\,I~6393\,\AA{}  absorptions in the spectra taken on different dates are compared, 
apart from line doubling, line depth variability is clearly seen.

\begin{figure} 
\hspace{-2cm}
\hbox{
\includegraphics[angle=0,width=0.39\columnwidth,height=0.6\columnwidth,bb=20 30 570 780,clip]{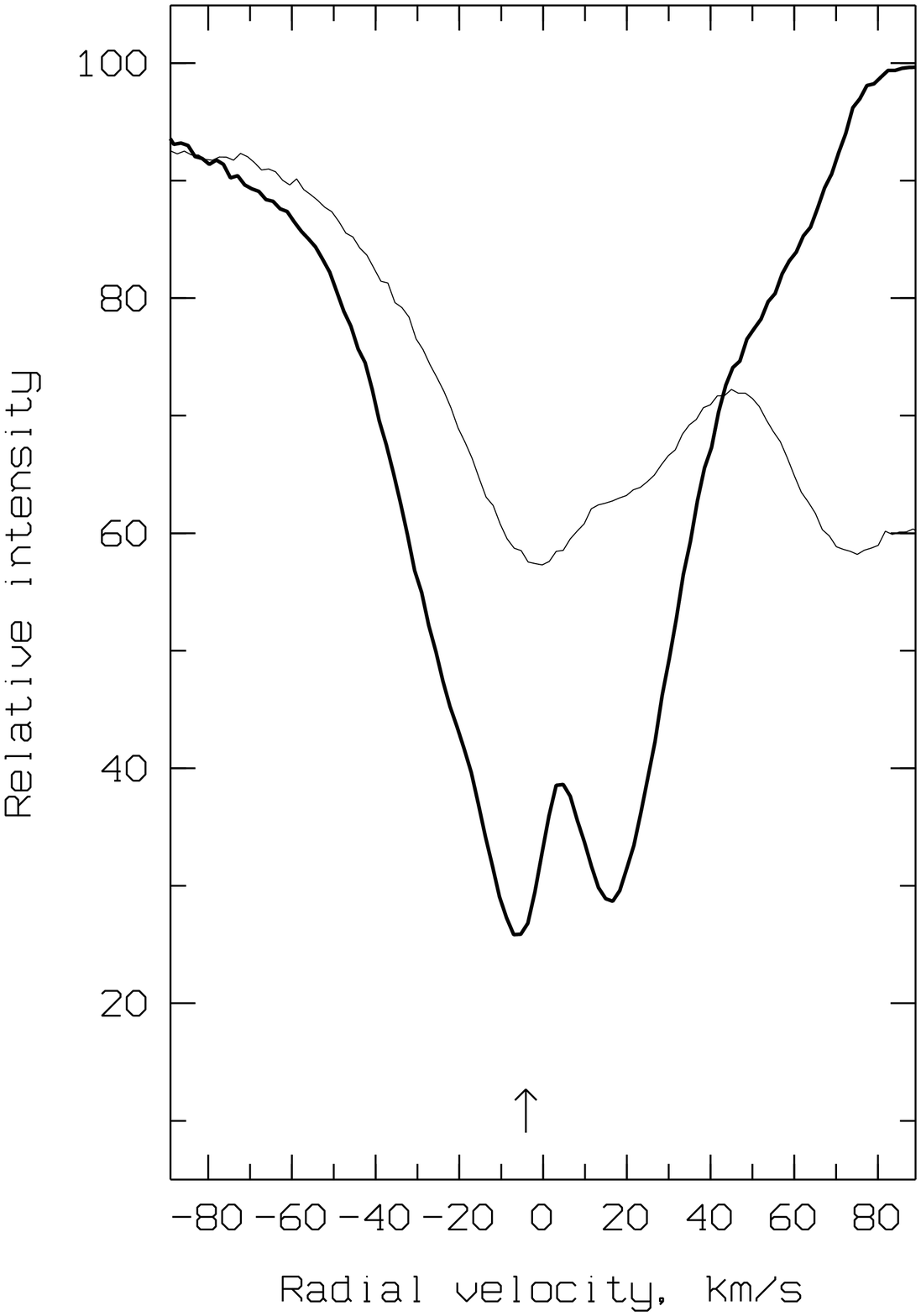}
\includegraphics[angle=0,width=0.39\columnwidth,height=0.6\columnwidth,bb=20 30 570 780,clip]{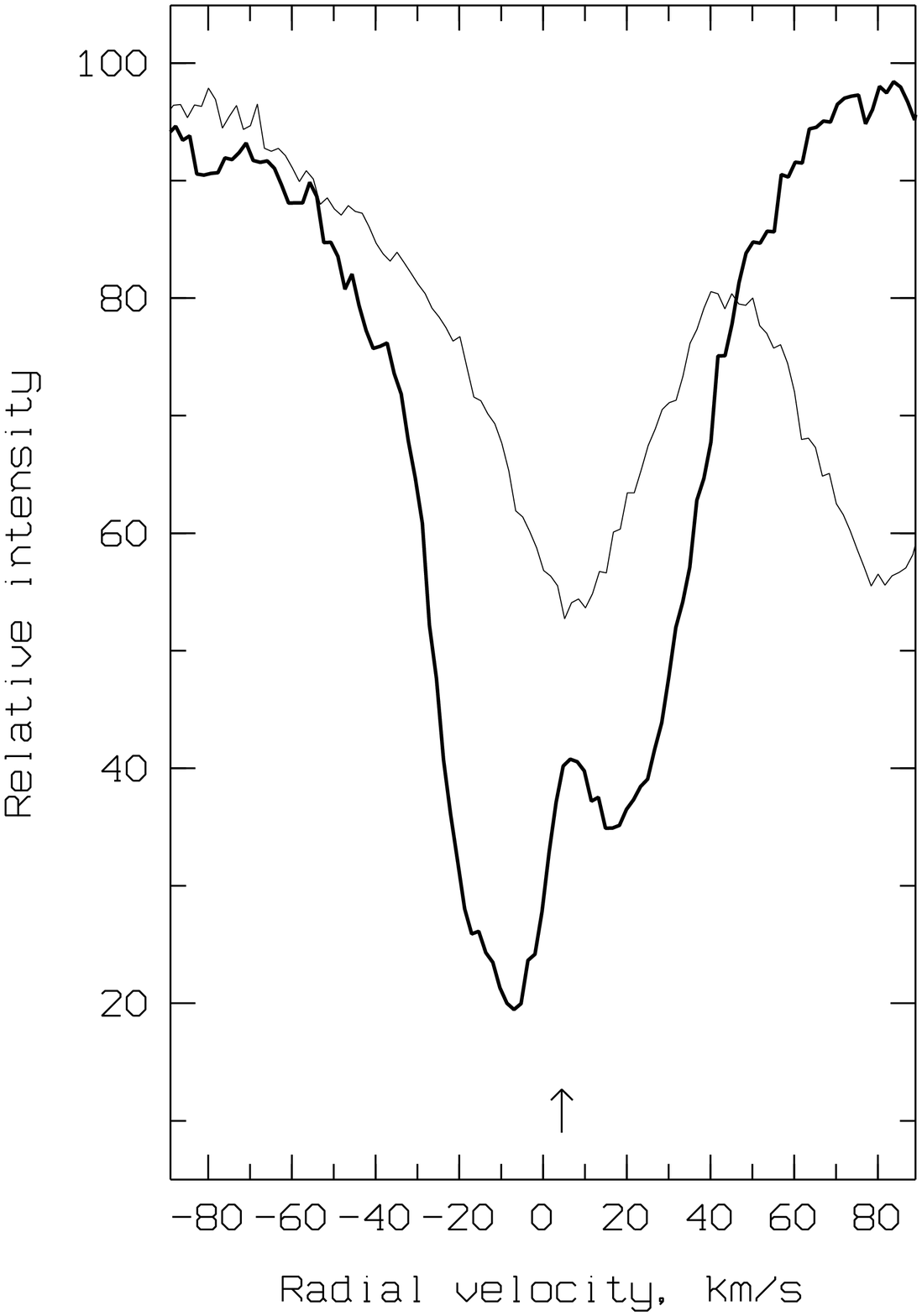}
\includegraphics[angle=0,width=0.39\columnwidth,height=0.6\columnwidth,bb=20 30 570 780,clip]{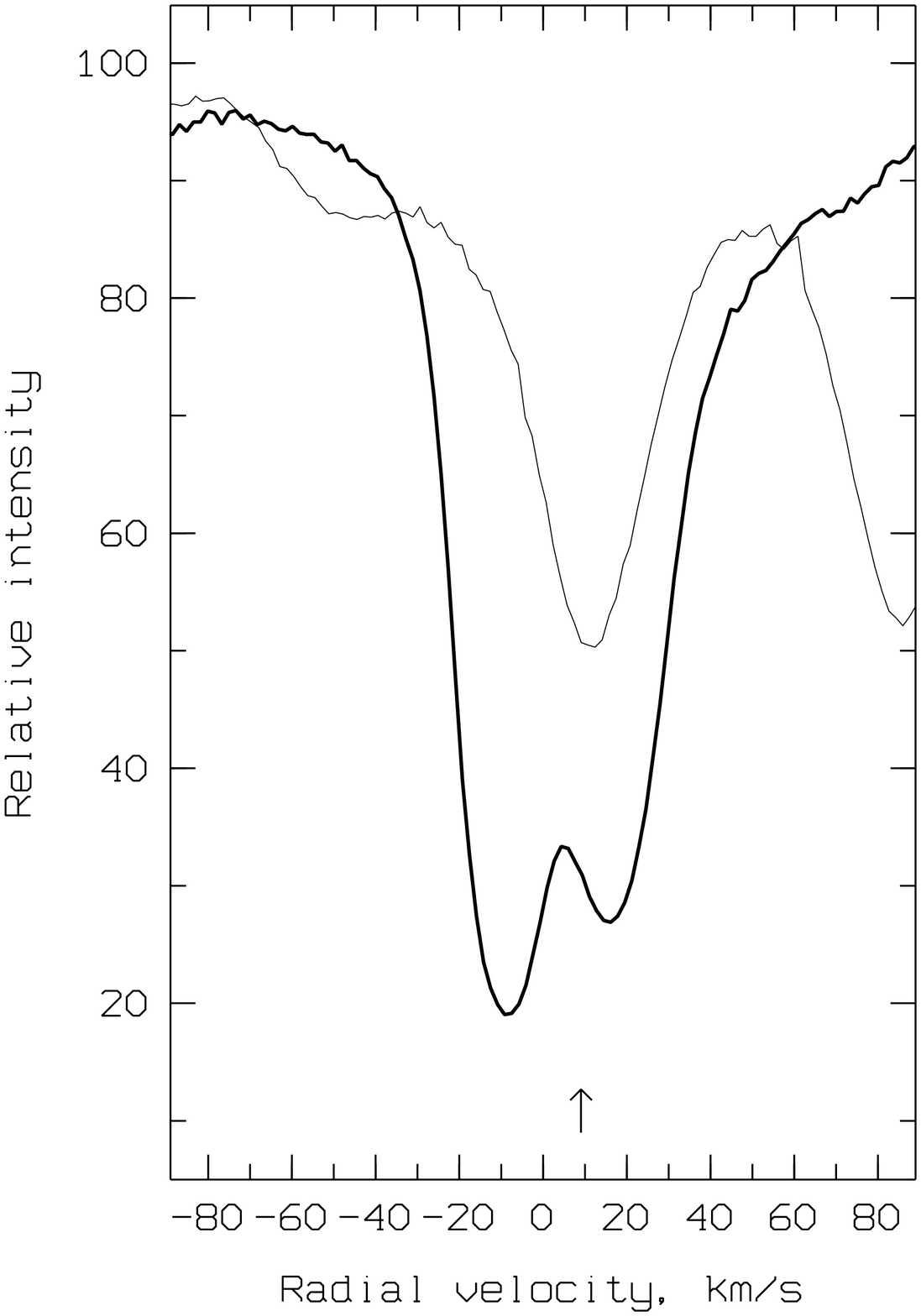}}
\caption{Fe\,II~6147\,\AA{} (thin curve) and Ba\,II~6141\,\AA{} (thick solid curve) line profiles in the October~13, 2013, April~28, 2015, and
        September~3, 2015 spectra of V1027 Cyg. The arrows indicate the mean radial velocities Vr(abs) from Table\,\ref{Radvel} measured from 
        symmetric absorptions  in the spectra for each time of observations.}
\label{ProfilesVar}  
\end{figure}

The difference in the behavior of lines with different intensities is shown in Fig.\,\ref{ProfilesVar}, where the profiles of 
the medium--intensity Fe\,II~6147\,\AA{}  line are compared with the Ba\,II~6141\,\AA{} line profiles. As can be seen from 
Figs.\,\ref{BaFe} and \ref{ProfilesVar},  the blue wing in the short--wavelength component of the Ba\,II~6141 line is steeper than 
the red wing of the long--wavelength component. This difference may point to a difference between the physical conditions under which 
these spectral features are formed. The shape of the long-wavelength absorption profile is more stable, while the shape of the blue wing in the
short-wavelength absorption changes significantly with time. The blue wing of the strong absorptions can be assumed to be formed under 
nstable conditions in the outer layers of the supergiant’s extended atmosphere affected by pulsations. Note also that the blue wing 
in the short-wavelength absorptions changes synchronously with the H$\alpha$ core.

It is important to emphasize in connection with the revealed anomalies of the strong absorptions that the abundances of chemical 
elements in the atmosphere of V1027\,Cyg calculated previously by Klochkova et al.~(2000a) do not require any correction, because 
in their calculations these authors used lines of a limited intensity with equivalent widths smaller than 200\,m\AA{}.

The only emission whose intensity exceeds considerably the continuum level in the spectrum of V1027\,Cyg is the emission component of the 
K\,I~7696\,\AA{} line. The P\,Cyg profile of this line in ``Relative  intensity -- Radial velocity'' coordinates is presented in Fig.\,\ref{K7696} 
in comparison with the profile of the resonance Na\,I~D lines. Comparison of the K\,I~7696\,\AA{} and Na\,I~D line profiles allows us to suspect 
the presence of a weak emission feature in the long--wavelength wing of the Na\,I~D lines.

\subsection*{Kinematic state of the stellar atmosphere}

The presence of peculiarities in the profiles of strong lines and their variability suggest that differential line shifts are probable 
in the spectrum and that these shifts are variable in time. Therefore, for our search of radial velocity variations we selected
low- and moderate-intensity absorptions with symmetric profiles without any visible peculiarities. Such absorptions are formed in deep 
layers of the stellar atmosphere and are not subjected to the possible influence of kinematic effects in the uppermost layers of the 
extended atmosphere. Tables\,\ref{Spectra} and \ref{Radvel} present the heliocentric radial velocities Vr measured from various types 
of spectral features  in the available spectra of V1027\,Cyg. We will rely on these data in our subsequent analysis of the spectrum peculiarities.

The mean velocities Vr(abs) corresponding to the positions of extensive samples of symmetric absorptions are given in column~4 of Table\,\ref{Spectra}. 
For three dates of observations we obtained radial velocities in the range from $-4.1\pm 0.1$ to $+9.1\pm 0.1$\,km/s. This variability of 
Vr(abs) is a manifestation of the pulsations of the V1027\,Cyg atmosphere. Recall that the temperature of V1027 Cyg is Teff\,=\,5000\,K, which, 
according to Soker~(2008), is close to the boundary of the transition from the AGB stage to the succeeding post-AGB one. As follows from the paper
by Aikawa (2010), the AGB star with parameters  Teff\,=\,5000\,K and log\,g\,=\,1.0 entered the instability strip
and can have radial pulsations. Having analyzed the long-term behavior of the UBV data for V1027\,Cyg, Arkhipova et al.~(2000) reached 
the conclusion about pulsations of the star in the fundamental mode and the first overtone with periods of P$\approx$\,306 and 250~days.

The results of our radial velocity measurements from the position of the component in the strongest absorptions are quite unexpected. 
The mean velocities from the full profiles of these strong features Vr(full) are given in the last column of Table\,\ref{Radvel}. The accuracy of 
measuring the position of the full profile for the split absorptions is lower than that for the remaining features of their profile 
due to its asymmetry. However, the accuracy is sufficient to conclude that the positions of the profiles are constant. These positions 
of the strong absorptions are clearly seen in Fig. 4, where the profiles of the strong Ba\,II~6141\,\AA{} line and the moderate--intensity 
Fe\,II~6147\,\AA{} line for three times of observations are compared on separate panels. Here, the position of the weak Fe\,II~6147\,\AA{} 
line in each spectrum coincides with the mean radial velocity inferred from symmetric absorptions Vr(abs).
At the same time, the position of the full Ba\,II~6141\,\AA{}  profile barely changes from date to date and does not correspond to Vr(abs). 
This behavior can be explained by the formation of this line in the outer atmospheric layers unaffected by pulsations.
For the strong absorptions we additionally measured the positions of the short-wavelength, Vr(blue), and long-wavelength, Vr(red), 
components as well as the position of the emission peak between them Vr(emis). The results of our Vr measurements from these components 
and their errors are given in Table\,\ref{Radvel}, columns 2--4, whence we conclude that the radial velocities in columns 2--4 measured with a high 
accuracy are constant with time. It follows from the constant positions of the emission features that they are formed in the envelope regions 
that are external with respect to the supergiant atmosphere.

\begin{table} 
\caption{Mean heliocentric radial velocities Vr measured from the features of strong absorptions: from the cores of the
        short-wavelength, Vr(blue), and long-wavelength, Vr(red), components, from the emission peak Vr(emis), and from the
        full profile Vr(full). The number of measured features is given in parentheses} 
\begin{tabular}{r|  c | c |  c| l}    
\hline
 \small Date &\multicolumn{4}{c}{\small  Vr, km$/$s} \\ 
\cline{2-5}  
          & \small blue  &\small emis & \small red & \hspace{0.5cm} full  \\ 
\hline               
13.10.2013&$-5.2\pm$0.2\,(63)&4.6$\pm$0.1\,(82)     &$16.0 \pm0.1$\,(83)  &$6.2 \pm0.3$\,(26) \\  
28.04.2015&$-5.0\pm$0.3\,(28)&6.1$\pm$0.1\,(15)     &$15.5 \pm0.2$\,(17)  &$3.3 \pm0.5$\,(7) \\   
 3.09.2015&$-5.1\pm$0.3\,(28)&5.4$\pm$0.2\,(33)     &$16.4 \pm0.2$\,(29)  &$4.7 \pm0.3$\,(27) \\  
\hline
\end{tabular}
\label{Radvel}
\end{table}

We have already noted above that the neutral hydrogen line profile is an absorption without any clear peculiarities expected for a G-type supergiant. 
However, as follows from Table\,\ref{Radvel}, the position of the H$\alpha$ core systematically differs from the expected one corresponding to the mean 
radial velocity Vr(abs) for the corresponding date. A possible explanation of the H$\alpha$ position in the spectrum is suggested by Fig.\,\ref{Halpha-3dat}, 
where an asymmetry of the line core apparently attributable to the influence of an aspherical envelope is clearly seen for all three times of observations. 
Evidence for asphericity of the V1027\,Cyg envelope also follows from the high degree of polarization  (P\,=\,7.7\%) that was determined by Trammell et al.~(1994) 
from spectropolarimetric observations. Subsequently, having performed multicolor polarimetric photometry, Parthasarathy et al.~(2006) obtained a degree of polarization 
$\approx 1$\%. 

In the far red region of the spectrum for V1027\,Cyg we identified  CN molecular lines. A fragment of the spectrum in the region 7960--8000\,\AA{}  
containing many weak lines (with depths 10--15\% below the continuum level) is presented in Fig. 6. We took the wavelengths for the CN features 
from the VALD database. The coincidence of the radial velocities measured from symmetric metal absorptions and CN lines points to the formation 
of the CN spectrum in the atmosphere of V1027\,Cyg, which is natural for such a cool star.

\subsection*{Interstellar Features and the Systemic Velocity Problem}

Apart from the numerous absorptions forming in the supergiant atmosphere, we detected at least ten weak absorptions in each spectrum of V1027\,Cyg
whose positions allowed them to be identified with DIBs. Table\,\ref{Spectra} gives the mean velocities corresponding to the positions of these features. 
Given the results of Vallee~(2008), the velocity averaged over three spectra, Vr(DIBs)\,=$-12.0$\,km/s, corresponds to the velocity of the 
interstellar medium in the Local Arm of the Galaxy.

\begin{figure}[ht!] 
\includegraphics[angle=0,width=0.6\columnwidth,height=0.8\columnwidth,bb=20 30 570 780,clip]{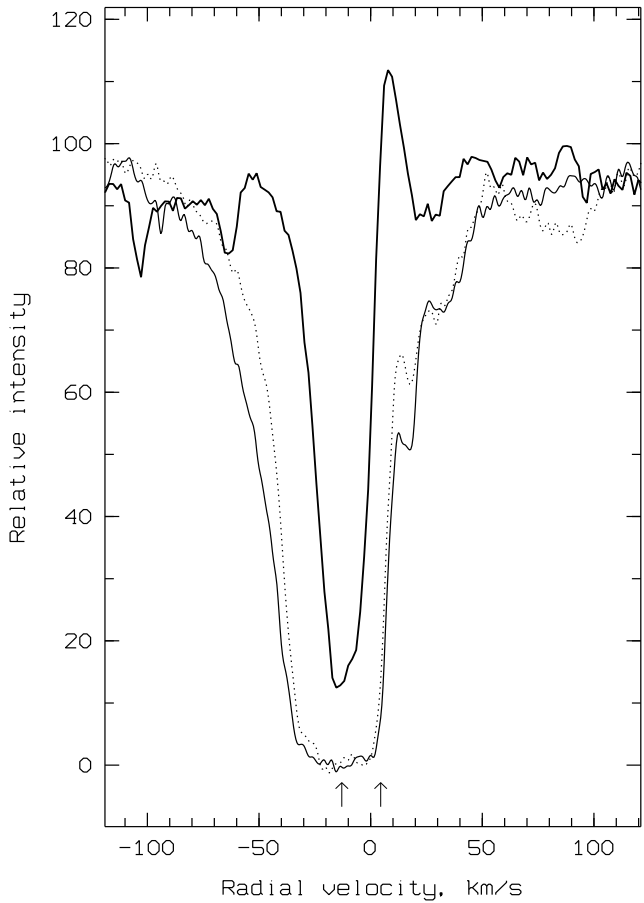}   %
\caption{Na\,I~5889\,\AA{} (thin curve), 5895\,\AA{}    (dotted curve), and K\,I~7696\,\AA{} (thick solid curve) 
        line profiles in the April~28, 2015 spectrum of V1027\,Cyg. The arrows indicate the velocities in the spectrum 
        for this date from Table\,\ref{Radvel} estimated from symmetric photospheric absorptions and DIBs: Vr(abs)\,=\,4.5\,km/s and 
        Vr(DIB)=$-12.9$\,km/s, respectively.}
\label{K7696}
\end{figure}

Consider in more detail Fig. 5, where the Na\,I and K\,I~7696\,\AA{}   line profiles are compared and the mean velocities inferred from numerous
symmetric absorptions, Vr(abs)\,=\,4.5\,km/s, and DIBs,  Vr(DIB)\,=$-12.9$\,km/s, for a given date of observations are indicated. 
Here, it can be seen that apart from the photospheric and interstellar (Vr(IS)\,=$-12.0$\,km/s) components, the broad NaI\,D line profile 
apparently contains the circumstellar and additional interstellar components (Vr$\le -17.0$\,km/s) that are not separated at the spectral 
resolution of the NES spectrograph. The interstellar components of these lines merge together to form a broad absorption in the velocity range 
from Vr$\approx -17.0$ to Vr$\approx -33.0$\,km/s. The longestwavelength component with Vr\,$\approx -12.0$\,km/s, just as the DIBs, 
is formed in the Local Arm of the Galaxy (Vallee 2006). As follows from Fig.\,\ref{K7696} and the data in the last column of Table\,\ref{Spectra}, 
the position of the absorption component of the K\,I~7696\,\AA{}  line with Vr$\approx -12$\,km/s is consistent with the DIB positions. 
This coincidence may point to an interstellar formation of the K\,I absorption component.
The presence of a short--wavelength component with Vr$\approx -33.0$\,km/s, which, according to Vallee~(2006), corresponds to the velocities 
in the Perseus arm, points to a more distant position of the star. We emphasize that Kipper and Klochkova~(2006) revealed the same interstellar 
velocity components (Vr$\approx -12.0$ and $-33.0$\,km/s) in the spectrum of the faint star that is the optical counterpart of the IR 
source IRAS\,20000+3239 with Galactic coordinates close to those of  V1027\,Cyg: ${\rm l/b}\approx 69/01$ and $67/00$ for IRAS\,20000+3239 and
V1027\,Cyg, respectively.

The absence of information about the star systemic velocity prevents us from unambiguously interpreting the pattern of radial velocities 
in the atmosphere and envelope of V1027\,Cyg. There is no information about the detection of molecular or maser emission in the published 
data on radio spectroscopy for IRAS\,20004+2955 (Eder et al. 1987; Lewis et al 1987; Bujarrabal et al. 1992). In the
absence of such data, as the systemic velocity of V1027\,Cyg we may take Vsys$\approx 5.5$\,km/s, which is the value of Vr 
averaged over all the emissions we detected: in the cores of the strong metal lines from Table\,\ref{Radvel} and Ve(7696)\,=\,6.1\,km/s 
from the emission in the K\,I~7696\,\AA{} line.

All of the available Vr measurements based on the atmospheric absorptions from Table\,\ref{Radvel} with those published previously 
by Klochkova et al.~(2000a) fall within the narrow range $-4.1\div 10.3$\,km/s. The adopted systemic velocity Vsys\,=\,5.5\,km/s leads 
to the tentative conclusion about a small radial velocity amplitude attributable to pulsations. Recall that Arkhipova et al.~(1997) 
obtained a wider range of values, Vr\,=$-10\div +20$\,km/s, from low-resolution spectra.

\section{Discussion}

Anomalous profiles of strong absorptions (Si\,II, Ni\,I, Ti\,I, Ti\,II, Sc\,II, Cr\,I, Fe\,I, Fe\,II, Ba\,II) have been detected in the spectrum 
of V1027 Cyg for the first time. Our measurements of the positions of individual features in these absorptions allow the profiles of these 
lines to be considered as a single broad absorption whose position does not correspond to Vr(abs) and does not change from date to date. 
This discrepancy between the velocities inferred from strong and weak absorptions can be explained by the fact that the strong absorptions 
are formed in the outer layers of the extended atmosphere unaffected by pulsations. Stable weak emissions are observed in the cores of all 
the strongest absorptions. The mean velocity of these emissions, in the absence of radio spectroscopic data for the IR source IRAS\,20004+2955,
may be considered as the systemic velocity Vsys\,=\,5.5\,km/s. 

Similar peculiarities of the spectrum are observed in R\,Sct, an RV\,Tau variable star, whose evolutionary status is close to the evolutionary 
stage of V1027\,Cyg. For example, Kipper and Klochkova~(2013) pointed out a splitting of the strongest absorptions in the spectrum of R\,Sct 
and the presence of a weak and variable emission in Fe\,I and Ti\,I lines.
However, as has been noted above, in contrast to V1027\,Cyg, the spectrum of R\,Sct near minimum light exhibits an emission in H$\alpha$ and H$\beta$ as well.

\begin{figure}[ht!] 
\includegraphics[angle=0,width=0.9\columnwidth,height=0.6\columnwidth,bb=20 30 570 780,clip]{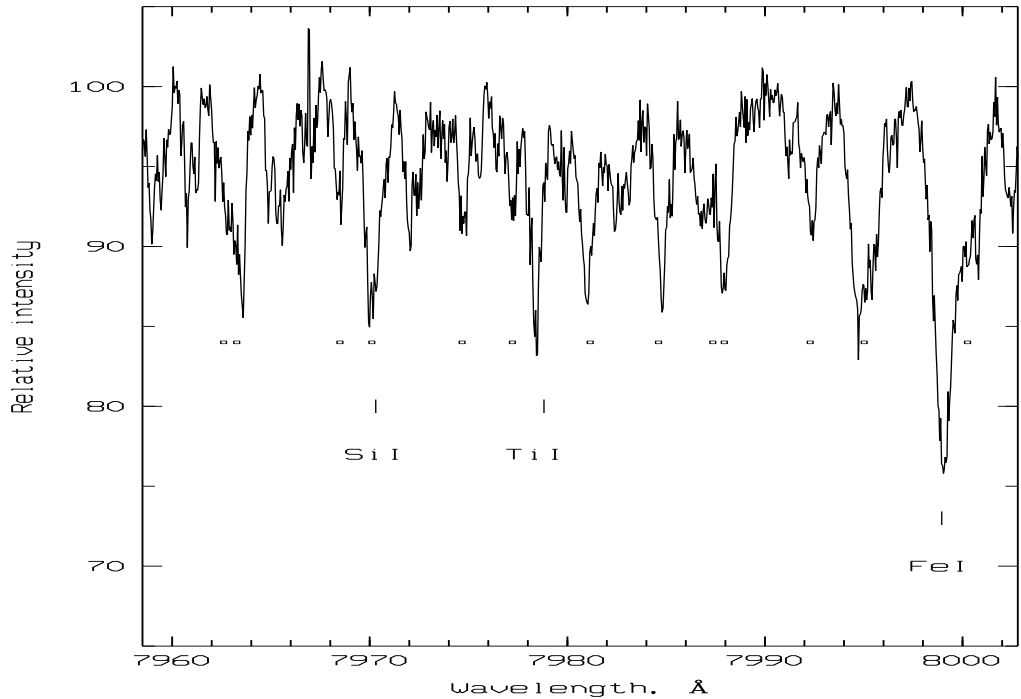}   %
\caption{Fragment of the spectrum for V1027\,Cyg. The dots mark the CN lines. The identified strongest (within this fragment) 
         metal absorptions are also indicated.}
\label{CN}
\end{figure}

Line doubling is a fairly common phenomenon observed in the spectra of variable stars of various types. The propagation of a shock wave is a universally 
accepted cause of the double absorption lines in the spectra of pulsating stars (for the details of this Schwarzschild scenario, see Alvarez et al.~2000).
The line doubling in the spectra for an extensive sample of Mira variables was investigated by Alvarez et al.~(2001). One of the most important 
results of their publication is that there is no visual-amplitude threshold beyond which line doubling only would occur in small-amplitude stars. 

Recall that doubling or asymmetry of strong absorptions were detected in the spectra of a small group of selected post-AGB stars. 
In contrast to V1027\,Cyg, the atmospheres of all stars from this sample are enriched with carbon and heavy metals. In particular, 
through a spectroscopic monitoring with a high spectral resolution Klochkova~(2013) detected
a splitting of strong absorptions with a low lowerlevel excitation potential in the spectrum of the post-AGB supergiant V5112\,Sgr 
for the first time. Invoking the radio spectroscopic data for the associated IR source IRAS\,19500$-$1709 led to the conclusion that 
the short-wavelength components of the split absorptions are formed in a structured circumstellar envelope. Thus, evidence for an 
efficient dredge-up of the heavy metals produced during the preceding evolution of this star into the envelope was obtained.
A similar effect was detected (Klochkova et al.~2015) in the spectrum of the optically faint star associated with the IR source IRAS\,23304+6147, 
with large overabundances of carbon and heavy metals having been revealed in its atmosphere (Klochkova et al.~2000b). The general properties 
of a sample of post-AGB supergiants with a splitting of strong absorptions are summarized in Klochkova and Panchuk~(2016), 
where it was shown that the type of profiles for strong absorptions could be connected with the morphology of the envelope and 
with its kinematic and chemical properties. 

However, despite the fact that the anomalies of the profiles of strong absorptions  in the spectra of the mentioned post-AGB stars 
and V1027\,Cyg are similar in appearance, we reached the tentative conclusion 
that the causes of the profile distortion in the spectra of these objects are different. In the spectra of post-AGB stars at least 
one of the components of the split absorptions is formed in the envelope detached from the central star. 
In contrast, in the case of V1027\,Cyg, the strong absorptions with split cores are formed in the upper layers of the extended stellar atmosphere.
Their positions are, on the whole, stable; the pulsations manifest themselves as a weak variability of the short-wavelength wings. 
Thus, the kinematics in the atmosphere of V1027\,Cyg can be said to be stratified: small-amplitude pulsations are observed 
in deep layers of the stellar atmosphere, while the upper atmospheric layers are stable. Using their IR observations with the VLT, 
Lagadec et al.~(2011) studied the images for a large sample of supergiants and showed the image of V1027\,Cyg to be point-like; 
the circumstellar dust envelope is not resolved.

The DIB positions, Vr(DIBs)\,=$-12.0$\,km/s, correspond to the velocity of the interstellar medium in the Local Arm of 
the Galaxy (Georgelin and Georgelin 1970). Using the adopted systemic velocity Vsys$\approx 5.5$\,km/s (LSR$\approx 23$\,km/s) 
and the data from Brand and Blitz~(1993) on the velocity field in the Galaxy, we can estimate the distance to the star, d\,=\,3.9\,kpc. 
This estimate agrees well with the distance to V1027\,Cyg that Bogdanov and Taranova~(2009) obtained by an independent method.

The IR spectrum of V1027\,Cyg\,=\,IRAS\,20004+2955 exhibits a strong emission at 10\,$\mu$ identified with silicates in the stellar 
envelope. This emission, which is typical of the spectra of O-rich stars, served to Hrivnak et al.~(1989) as a basis to consider 
IRAS\,20004+2955 to be an analog of the well-studied source IRAS\,18095+2704. However, the central stars of these IR sources differ in metallicity. 
For the central star of IRAS\,18095+2704 ${\rm [Fe/H]}$\,=$-0.8\div -0.9$, according to Klochkova (1995) and Sahin et al.~(2011), respectively, while 
in the atmosphere of V1027\,Cyg, as Klochkova et al.~(2000a) showed, the metallicity is nearly solar: ${\rm [Fe/H]}$\,=$-0.2$. 
The high metallicity of V1027\,Cyg and its location close to the Galactic plane suggest that this star belongs to the disk population. 

We think the central star of the IR source IRAS\,20508+2011 already mentioned in the text to be a closer relative for V1027\,Cyg. 
The parameters of this star derived by Klochkova et  al.~(2006) (its luminosity Mv$\approx -3$\,mag, effective temperature
Teff\,=\,4800\,K, surface gravity log\,g\,=\,1.5, metallicity ${\rm [Fe/H]}$\,=$-0.36$, and chemical composition in whole) are typical
of an AGB star. The ratio of the carbon and oxygen overabundances allows  IRAS\,20508+2011 to be assigned to the group of evolved 
stars with oxygen enriched atmospheres (${\rm [O/C]}$\,=+0.9), which is consistent with the object position on the IR color-color diagram. 
No overabundance of s-process elements was detected, just as in V1027\,Cyg. Since as yet there are no observations in water 
maser and OH bands for IRAS\,20508+2011, we cannot unambiguously classify the envelope. However, it can be asserted that in 
the case of IRAS\,20508+2011 we observe an extremely early PPN formation phase that comes immediately after the termination 
of mass loss and apparently until the beginning of envelope separation.

As has already been noted by Klochkova et al.~(2000a), the set of main parameters of V1027\,Cyg (the location of this star close 
to the Galactic plane, its nearly solar metallicity, and the atmospheric abundances of chemical elements) more likely corresponds 
to a classical supergiant, especially since the intensity of the O\,I~$\lambda 7771-7775$\,\AA{} triplet, which served as a universally 
accepted luminosity indicator, is very high in the spectrum of this star. The total equivalent
width of the triplet in the spectrum of V1027\,Cyg is W(7773)\,=\,1.86\,\AA{}. Applying the Mv--W(7773) calibrations published 
by Klochkova et al.~(2002) and  Arellano Ferro et al.~(2003), we obtain the luminosity Mv\,=$-8.0$\,mag for V1027\,Cyg, which is 
considerably higher than the luminosity of AGB stars but agrees well with luminosity class Ia. However, we should keep in mind 
the possible error in the luminosity due to the application of the Mv--W(7773) calibration obtained from Cepheids and other 
massive A--G supergiants of the Galactic disk to a star with an unclear evolutionary status and the neglect of possible 
differences in oxygen abundance.

Taking the luminosity of the star to be Mv\,=$-8.0$\,mag and the color excess to be E(B$-$V)\,=\,0.6\,mag  (Arkhipova et al.~2000), 
we estimate the distance to V1027\,Cyg to be d$\approx 7.5$\,kpc, which exceeds considerably the distance d\,=\,3.9 kpc corresponding 
to the systemic velocity of the star we adopted. The unreliability of the color excess and systemic velocity estimates is responsible for 
the discrepancy. 
For example, if we use the interstellar reddening Av\,=\,3.6\,mag estimated by Hrivnak et al.~(1989) for V1027\,Cyg by modeling the flux 
in the wide wavelength range 0.4--100\,$\mu$, then we will obtain d$\approx$4.0\,kpc. This value agrees well with the distance inferred 
from the systemic velocity.

To refine our conclusions about the star V1027\,Cyg, we need its spectroscopic monitoring with an ultrahigh spectral resolution (R$\ge 10^5$) 
that will ensure the separation of the components of the Na\,I~D and K\,I~7696\,\AA{} line profiles, the refinement of the proposed velocity 
pattern, and, consequently, the refinement of the distance to the star and its evolutionary status.

\section{Conclusions}

Based on our high-spectral-resolution observations performed with the NES echelle spectrograph of the 6-m telescope, 
we detected the peculiarities of the spectrum and the velocity field in the atmosphere of the cool supergiant V1027\,Cyg, 
the optical counterpart of the IR source IRAS\,20004+2955. Small variations of the radial velocity Vr(abs) with an amplitude of about 
5\,km/s due to pulsations were revealed by symmetric low- and moderate-intensity absorptions.

Numerous weak CN molecular lines and the K\,I~7696\,\AA{} line with a P\,Cyg profile were identified in the red spectral region. 
The coincidence of the radial velocities measured from symmetric metal absorptions and CN lines suggests that the CN spectrum 
is formed in the stellar atmosphere. We identified also numerous diffuse interstellar bands  whose positions in the spectrum, 
Vr(DIBs)\,=$-12.0$\,km/s, correspond to the velocity of the interstellar medium in the Local Arm of the Galaxy.

A long-wavelength and time-variable shift of the H$\alpha$ profile due to line core and short-wavelength wing distortion is observed 
in the spectrum of V1027\,Cyg. A splitting of the cores of the strongest absorptions of metals and their ions (Si\,II, Ni\,I, Ti\,I, 
Ti\,II, Sc\,II, Cr\,I, Fe\,I, Fe\,II, Ba\,II) has been detected in the stellar spectrum for the first time. The broad profile 
of these lines contains a stable weak emission in the core whose position may be considered as the systematic velocity Vsys\,=\,5.5\,km/s. 
Spectroscopy with an ultrahigh spectral resolution is needed for a more definite interpretation of the radial velocity pattern, 
for identifying the circumstellar and photospheric components in the Na\,I~D line profile, and, consequently, for refining the distance 
to the star and its evolutionary status.

\section*{Acknowledgments}
This work was supported by the Russian Foundation for Basic Research (projects no.~14-02-00291\,a and 16-02-587\,a).
We used the SIMBAD and ADS astronomical databases. We also used the VALD database maintained in the Uppsala and Vienna Universities
and at the Institute of Astronomy of the RAS (Moscow).

\end{document}